# A spatiotemporal adaptive local search method for tracking congestion propagation in dynamic networks

Weihua Huan[a, b], Kaizhen Tan[c], Xintao Liu[b], Shoujun Jia[d], Shijun Lu[e, a], Jing Zhang[f], Wei Huang [a, g, h, *]

[a] *College of Surveying and Geo-informatics, Tongji University, Shanghai 200092, China*

[b] *Department of Land Surveying and Geo-Informatics, The Hong Kong Polytechnic University, Kowloon, Hong Kong SAR, China*

[c] *School of Economics and Management, Tongji University, Shanghai 200092, China*

[d] *Department of Geography, University of Innsbruck, Innsbruck, Austria*

[e] *College of Civil Engineering and Architecture, Xinjiang University, Urumqi 830000, China*

[f] *School of Civil Engineering and Architecture, Xiamen University of Technology, Xiamen, Fujian 361024, China*

[g] *Department of Civil Engineering, Toronto Metropolitan University, Toronto, M5B 2K3, Canada*

[h] *Urban Mobility Institute, Tongji University, Shanghai 200092, China*

**Corresponding author:** Wei Huang, wei_huang@tongji.edu.cn

**Permanent address**: College of Surveying and Geo-informatics, Tongji University (1239 Siping Road, Yangpu District, Shanghai, 200092)

**Abstract**: Traffic congestion propagation poses significant challenges to urban sustainability, disrupting spatial accessibility. The cascading effect of traffic congestion propagation can cause large-scale disruptions to networks. Existing studies have laid a solid foundation for characterizing the cascading effects. However, they typically rely on predefined graph structures and lack adaptability to diverse data granularities. To address these limitations, we propose a spatiotemporal adaptive local search (STALS) method, which feeds the dynamically adaptive adjacency matrices into the local search algorithm to learn propagation rules. Specifically, the STALS is composed of two data-driven modules. One is a dynamic adjacency matrix learning module, which learns the spatiotemporal relationship from congestion graphs by fusing four node features. The other one is the local search module, which introduces local dominance to identify multi-scale congestion bottlenecks and search their propagation pathways. We test our method on the four benchmark networks with an average of 15,000 nodes. The STALS remains a Normalized Mutual Information (NMI) score at 0.97 and an average execution time of 27.66s, outperforming six state-of-the-art methods in robustness and efficiency. We also apply the STALS to three large-scale traffic networks in New York City, the

United States, Shanghai, China, and Urumqi, China. The ablation study reveals an average modularity of 0.78 across three cities, demonstrating the spatiotemporal-scale invariance of frequency-transformed features and the spatial heterogeneity of geometric topological features. By integrating dynamic graph learning with Geo-driven spatial analytics, STALS provides a scalable tool for congestion mitigation.

## 1. Introduction

Traffic congestion propagation in dynamic networks is characterized by spatiotemporally heterogeneous cascading effects, where localized disruptions propagate asymmetrically across the network through complex node-link interactions (Basak, Dubey, and Bruno 2019). This cascading effect of traffic congestion propagation could induce secondary and tertiary congestion, because it spreads dynamically through transportation networks from the congestion source, regardless of recurrent and non-recurrent congestion (Anbaroglu, Heydecker, and Cheng 2014). Particularly problematic include two aspects: (1) the emergence of the congestion bottleneck points where congestion accumulates and radiates, disproportionately affecting overall network functionality, and (2) time-varying propagation paths that render static analysis based on fixed network characteristics ineffective. The consequences of these propagation dynamics are severe, not only undermining human mobility efficiency but also hindering the development of a sustainable transportation system through excessive energy consumption and system-wide inefficiencies (Faheem, Shorbagy, and Gabr 2024; Ding et al. 2025). Therefore, the key to formulating effective congestion mitigation strategies lies in detecting the traffic congestion bottleneck points and modelling their propagation pathways.

Exensive efforts have been devoted to the research of traffic congestion propagation modelling (**Figure 1**), with simulation-based methods as the first branch. Microscopic simulation takes vehicles as a research object and transforms the congestion propagation into a dynamic queuing process, such as car-following (Zhang et al. 2024a) and lane-changing (Ma and Li 2023) models. For example, Klawtanong and Limkumnerd (2020) utilized a stochastic car-following model to study the dissipation rule of traffic congestion and proposed a safety distance for drivers. Madaan and Sharma (2022) considered both drivers' behavior and the lane-changing phenomenon to simulate traffic dynamics, revealing a nontrivial influence between them. While this approach provides a nuanced simulation of individual vehicle interactions, it is rarely available in large-scale networks, causing error accumulation in propagation prediction. On the contrary, macroscopic simulation focuses on the overall characteristics of traffic flow, represented by the LWR proposed by Lighthill and Whitham (1955) and Richards (1956), the cell transmission model (CTM), and susceptible-infectious-recovered (SIR) models. LWR describes road network behavior using fluid dynamics, widely applied to detecting the traffic congestion bottlenecks of urban highways. CTM is the most widely adopted and has evolved into many variants over the decades (Wu et al. 2022). For example, Rrecaj et al. (2021) and Peng et al. (2022) both uncovered the evolution mechanisms of traffic congestion based on the CTM. Wang et al. (2024) introduced the susceptible-infectious-suusceptible (SIS) and proposed The SIS-CTM to describe the time-verying traffic characteristics. While Wu, Gao, and Sun (2004) pioneered the SIR-based framework for congestion propagation, their work remained hypothetical. Subsequent studies by Saberi et al. (2020) and Kozhabek et al. (2024) empirically validated the approach, confirming its generality across multi-city verification. Despite the efforts, scholars have pointed out the shortcomings of this approach - the "distortion" problem caused by theoretical assumptions (Yang and Wang 2020). They rely on underlying traffic flow models and oversimplify the network, especially when a fixed cell length is required by CTM, making it difficult to adapt to urban roads with highly irregular geometric conditions.

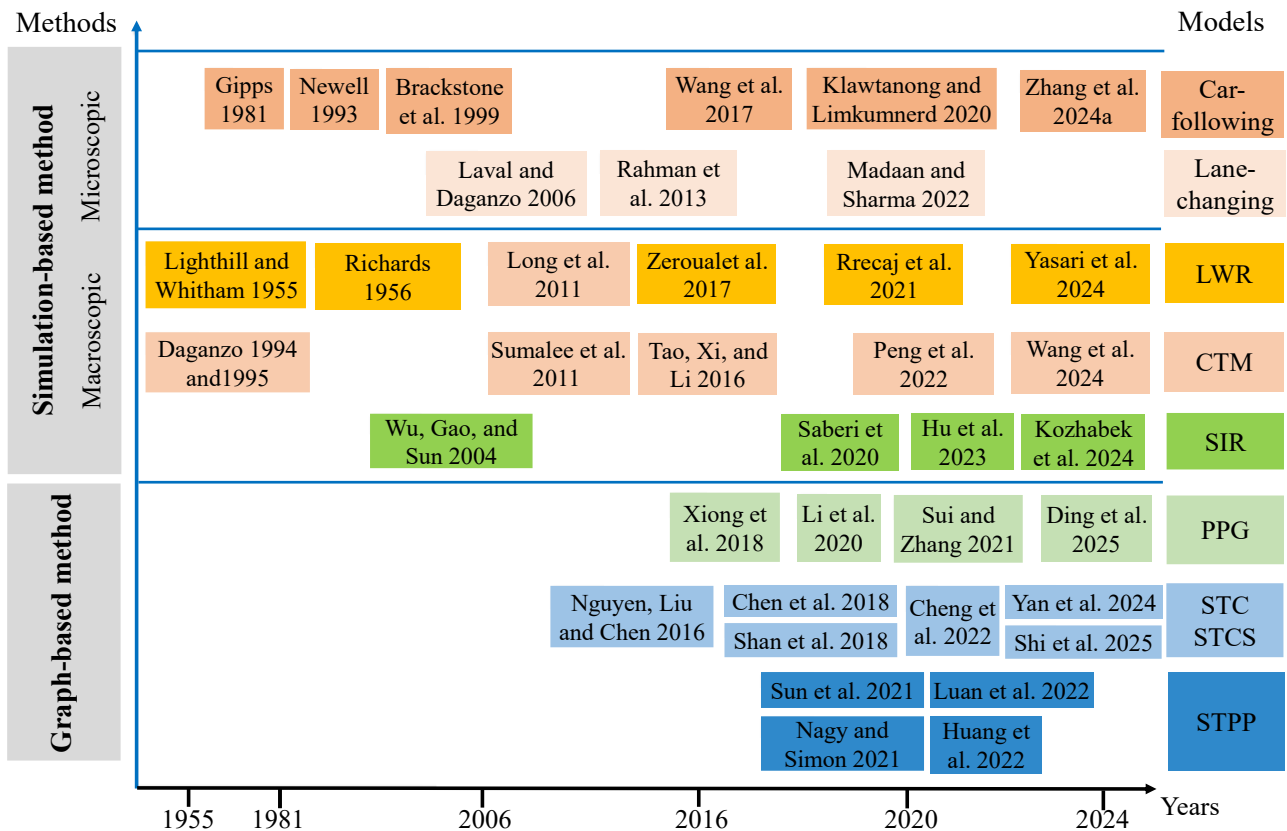

**Figure 1.** Overview of traffic ongetsion propagation analysis, represented by two branches: simulation-based and graph-based methods.

The second branch is graph-based methods, which model the road network as a graph without simplification. The propagation probability graph (PPG) method states that the congestion propagation has a Markov property (Xiong et al. 2018; Li et al. 2020; Sui and Zhang 2021; Ding et al. 2025), which means that the likelihood of a congestion propagating from road segment A to B is independent. PPG does not consider that the predecessor road segments may produce inprecise results. The spatiotemporal congestion graph (STC) method is a development of PPG. It aims to search congestion propagation by examining the congestion trees backward in time and finds the frequent congestion trees. For example, Nguyen, Liu and Chen (2016) built a probabilistic model with the frequent propagation trees using the dynamic Bayesian network (DBN) to determine the probability of a complete propagation pattern and reveal some causal interactions. As an extension of STC, Chen et al. (2018) further developed the spaiotemporal congestion subgraph (STCS) method to search for frequent congestion propagation trees by using the propagation path exploration of the STC. Spatiotemporal congestion propagation patterns (STPP)-based methods (Nagy and Simon

2021) reveal congestion propagation based on STC and STCS methods, receiving more and more attention in recent years because STPP could maintain a directed graph in which it stores possible propagation paths. For example, Sun et al. (2021) learned the behavior of traffic ongetsion propagation by STPP prediction based on low dimension embeddings of the road segments. Luan et al., (2022) and Huang et al. (2022) integrated graph and deep learning in STPP modelling for traffic predictione tasks, however, these graph neural network methods require ground truth labels to compare their performance with the state-of-the-arts. Scholars have emphisized that investigating the congestion cluster dynamics in evolving networks is crucial to reveal the hidden information during congestion propagation and dissapartion phases (Saeedmanesh and Geroliminis 2017). In this scope, the above works typically focus on traffic congestion propagation modelling based on network structure (e.g., propagation tree or graph) by mining frequent congested substructure from historical satasets. However, diritical chanlleges remain: (i) the definition of static adgacency relationship cannot fully capture the information-passing as the congestion graph evolves over time, and (ii) the temporal resolution of data (e.g., 5min and 1h) highly affects the outputs, as these methods reveal large-scale congestion dynamics through obscuring small-scale interactions (Xiong, Zhou, and Bennett 2023).

The above challenges prompt the proposal of a spatiotemporal adaptive local search (STALS) method, which learns congestion propagation rules by searching multi-scale communities from dynamically updated adjacency matrices. The proposed method consists of two data-driven modules - a dynamic adjacency matrix learning module for multi-feature fusion and a local search module for community detection. Experiments on four benchmark datasets demonstrate that the STALS outperforms the six competing state-of-the-art models. The contribution lies in three folds:

- A dynamic adjacency matrix adaptive to data at different time granularities is designed to capture the spatiotemporal interactions in evolving congestion graphs. Such matrices are derived by fusing four graph features, i.e., curvature, degree, spatial proximity, and semantics, instead of simple network connectivity.

- The proposed method searches for multi-scale communities with local dominance from those dynamic adaptive adjacency matrices, thus identifying multi-level congestion bottlenecks.
- A comparative analysis of the three large-scale traffic networks validates the ability of our proposed method to identify congestion bottlenecks and their propagation pathways at the road-segment level.

The remainder of this article is organized as follows. **Section 2** introduces the details of the proposed spatiotemporal adaptive local search method**. Section 3** validates the superiority of the STALS through benchmark experiments and demonstrates its performance using three real-world traffic datasets. **Section 4** discusses the STALS' significance and limitations. Finally, the conclusions are presented in **Section 5**.

## 2. Methodology

### *2.1. Preliminary definitions*

#### *2.1.1. Spatiotemporal congestion instance*

Traffic State Index (TSI) (Chen et al. 2018) is adopted to measure the traffic state in this study. Given a road segment $f_i$ at the timestamp $t_j$, the TSI is calculated by the relative deviation rate of actual speed to the free-flow speed:

$$TSI_{f_i,t_j} = \frac{v_{f_i} - \bar{v}_{f_i,t_j}}{v_{f_i}}, \tag{1}$$

where $v_{f_i}$ represents the free-flow speed of the road segment $f_i$, $\bar{v}_{f_i,t_j}$ is the actual average speed at the timestamp $t_j$. The range of $TSI_{f_i,t_j}$ is [0,1], and the threshold is 0.7. Therefore, the road segment $f_i$ at the timestamp $t_j$ is defined as a spatiotemporal congestion instance if its $TSI_{f_i,t_j}$ is no less than 0.7.

#### *2.1.2. Spatiotemporal congestion subgraph*

An undirected spatiotemporal congestion subgraph $G_{t_j}(V_{t_j}, E_{t_j}, W_{t_j})$ is defined as a set of all spatiotemporal congestion instances at the timestamp $t_j$, where $V_{t_j}, E_{t_j}, W_{t_j}$ represent nodes, edges, and weights, respectively. The nodes in the subgraph represent congested road segments, and the weights of edges are determined by the spatiotemporal relationship between adjacency nodes. A set of spatiotemporal congestion subgraphs for sequential periods is termed a spatiotemporal congestion graph:

$$G = \bigcup_{j=1}^{J} G_{t_j}(V_{t_j}, E_{t_j}, W_{t_j}), \qquad (2)$$

where $J$ is the total number of timestamps.

*2.1.3. Spatiotemporal adjacency relationship*

The edge values of $G_{t_j}(V_{t_j}, E_{t_j}, W_{t_j})$ are measured by the spatiotemporal relationship between two adjacency nodes. Although nodes are fixed by certain physical constraints, their relationship changes dynamically over time and evolves into spatiotemporal congestion subgraphs. In this study, we capture this relationship for each spatiotemporal congestion subgraph from four aspects: node geometry, node topology, spatial proximity, and semantic information.

- Node geometry. Existing studies have shown that curvature is a key structural property for characterizing large-scale networks and has a major impact on core congestion (Narayan and Aaniee 2011). Therefore, the curvature of each road segment is as the node geometry.

- Node topology. The degree of a node is an important indicator affecting information passing and the stability of the graph structure. It refers to the number of edges directly connected to the node. Nodes with higher degrees are considered to have higher centrality.

- Spatial proximity. Spatial proximity is used for measuring the spatiotemporal dependency and reflecting node interactions. According to Jiang and Luo (2022), spatial proximity mainly contains neighbor closeness and distance closeness. In this study, we choose distance closeness (i.e., Euclidean distance), since node degree has already captured neighbor proximity.

- Semantic information. Semantics offers additional dynamic knowledge beyond graph structure. The fast Fourier transform can extract the inherent nature of time series and eliminate the effects of temporal resolution. Inspired by this, a fast Fourier transform of the TSI matrix is adopted to capture dynamic traffic state information in this study.

## 2.2. STALS method

We propose a new data-driven method, namely STALS, for tracking congestion propagation in dynamic networks at the level of community structure. The STALS calculated dynamic adjacency matrices instead of predefined ones by fusing node curvature, node degree, spatial proximity, and TSI semantics. Two data-driven modules are included (**Figure 2**): the dynamic adjacency matrix learning module and the local search module for community detection. More specifically, the adaptive adjacency matrices designed from the spatiotemporal congestion graphs (Section 3.2.1) are fed into the local search algorithm (Section 3.2.2), where the multi-scale communities can be detected.

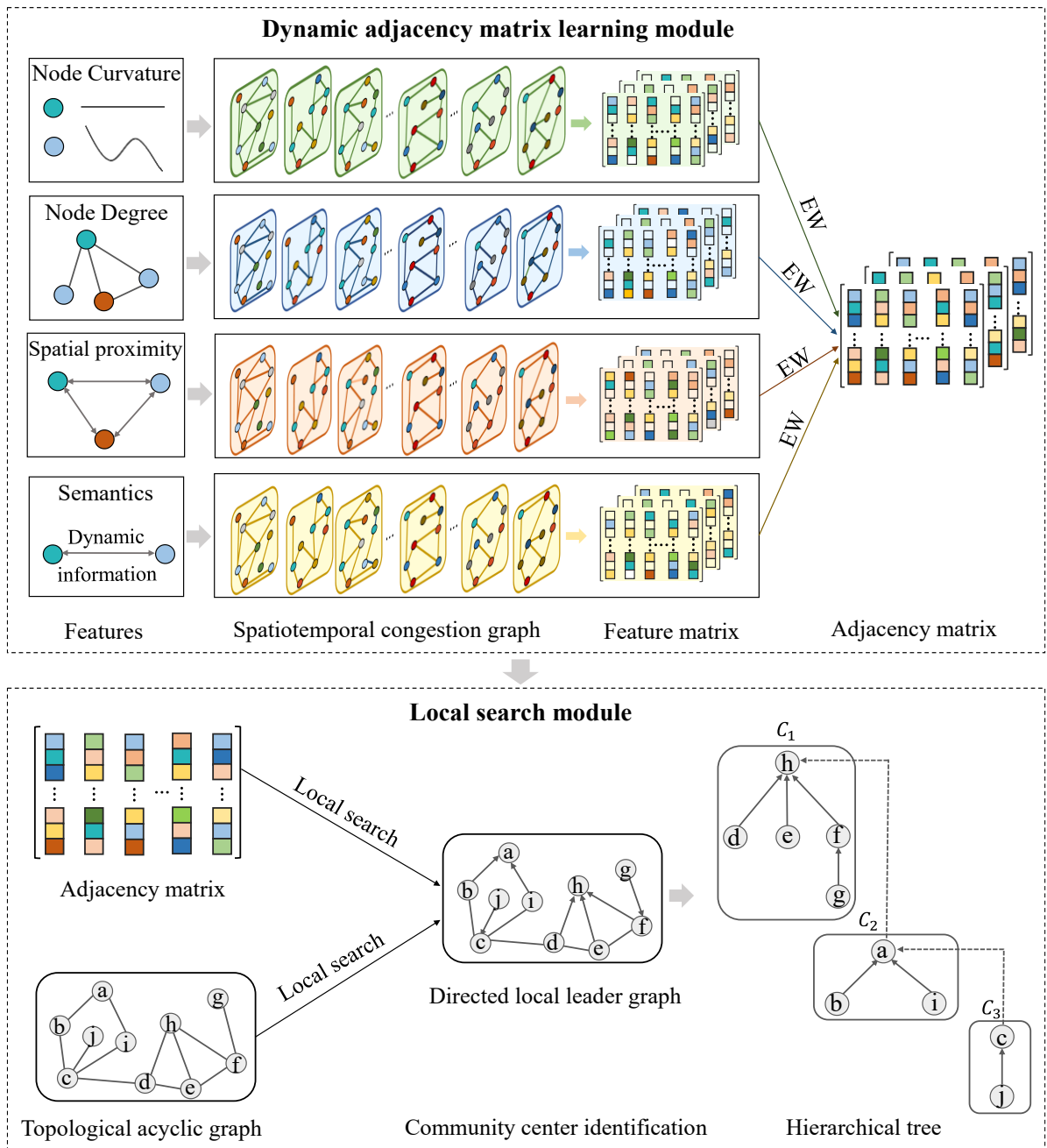

**Figure 2**. Framework of the proposed method.

*2.2.1. Spatiotemporal adjacency matrix*

Entropy weight (EW) is conducive to increasing the dipartite degree of relative closeness in multi-attribute decision-making (Chen 2021). Considering this merit, the EW approach is adopted to calculate adaptive adjacency matrices in this study. Given a spatiotemporal congestion subgraph $G_{t_j}$ at timestamp $t_j$, its adaptive adjacency matrix $M_{t_j}$ is calculated as the weighted fusion of its curvature similarity $K_{t_j}$, degree similarity $D_{t_j}$, Spatial proximity similarity $S_{t_j}$, and the Fourier transform of traffic state similarity $F_{t_j}$:

$$M_{t_j} = w_{t_j,1} K_{t_j} + w_{t_j,2} D_{t_j} + w_{t_j,3} S_{t_j} + w_{t_j,4} F_{t_j} \qquad (3)$$

where $w_{t_j,1}, w_{t_j,2}, w_{t_j,3}$, and $w_{t_j,4}$ represent the EWs calculated by the information entropy of each similarity matrix. $K_{t_j}$ and $D_{t_j}$ are calculated based on the Cosine similarity of node curvature and node degree, respectively. Given a spatiotemporal congestion graph $G$, its adjacency matrices at $J$ timestamps can be automatically updated as $M = [M_{t_1}, M_{t_2}, \ldots, M_{t_J}]$.

*2.2.2. Local search algorithm*

Based on the adaptive adjacency matrices, a local search algorithm identifies multi-scale congestion bottlenecks by analyzing leader-follower dynamics in traffic networks, The process consists of four key stages:

The first step is node value attribution, attributing a value $x_u$ to each node $u$ in the network. This is achieved by combining the adjacency matrix from the dynamic adjacency matrix learning module with the topological graph obtained from the spatiotemporal congestion graph. Specifically, $x_u$ is calculated by summing the values of all the edges that are directly adjacent to the node. This process enables the construction of a directed acyclic graph, where each node $u$ points to its largest-value-neighbor $v$ if the following conditions are met: (i) The node has a higher value than its neighbor, i.e., $x_u \leq x_v$, and (ii) each node u pointing to its largest-value-neighbor, i.e., $x_v = max\ \{x_i | i \in V(u)\}$, where $V(u)$ represents the set of all neighbors of $u$. The second stage is local leader identification (Blondel et al. 2008) based on the concept of local dominance. Local dominance refers to a leader-follower relationship in community structures, where nodes with incoming edge(s) but no outgoing edges are considered local leaders that dominate their surrounding area (see nodes a, c, and h in the module). These local leaders are crucial in understanding the network's hierarchical structure. Thirdly, for each local leader $u$, a local breath-first searching (LBFS) (Shi et al. 2024) algorithm is employed to find its nearest local leader $v$ with $x_u \leq x_v$. The advantage of the LBFS lies in its high efficiency, as it can terminate its search once the nearest local leader is identified, without traversing the entire network. It also allows for the determination of the shortest path length between local leaders, providing valuable insights into the network's connectivity. Finally, multi-scale communities denoted as the symbol C-$i$ in the following text can be identified, with $i$ symbolizing different levels and scales of communities. **Table 1** shows the pseudocode of the STALS. The complete data and code that support the findings of this study are available on Figshare (Huan, 2025).

**Table 1** Pseudocode of the STALS algorithm.

| **Algorithm**: Spatiotemporal adaptive local search algorithm flow |
|---|
| **Input**: Adjacency matrices at different timestamps |
| **Output**: The ID of identified communities, metrics |
| **function** calculate modularity (G, partition): |
|    communities = {} |
|    **for** node in G.nodes(): |
|       community = partition[node] |
|       **if** community **not in** communities: |
|          communities[community] = [] |
|      **append** node **to** communities[community] |
|    m = **total edges** (G) |
|    Q = 0.0 |
|    **for** community **in** communities: |
|       weight = **sum edge weights** (G, community) |
|       k = **size** (community) |
|       Q += (weight / m - (k / (2 * m))^2 / 2 |
|    **return** Q |
|    **for** time slice in range(T): |
|       // 1. Read adjacency matrix |
|       adjacency matrix = read_csv (feature_path[time slice]) |

```
    // 2. Build weighted graph
    G = create_graph ()
    for source, row in adjacency matrix:
       for target, weight in row:
          if weight > 0:
             add_edge (G, source, target, weight)

    // 3. Calculate node weights (congestion strength)
    for node in G.nodes ():
       weight = sum_weights (G, node)
       set_node_attribute (G, node, 'weights', weights)

    // 4. Extract largest connected component (LCC)
    largest_cc = find_largest_component(G)
    G_largest = subgraph(G, largest_cc)

    // 5. Detect congestion centers (optimize modularity)
    best_modularity = -1
    best_centers = []
    for leaders_num in [center_numbers]:
       D, centers, labels, partition = hierarchical_weight_communities(
          G_largest, leaders_num, auto_choose_centers=False, maximum_tree=True
       )
       modularity = calculate_modularity(G_largest, partition)

       if modularity > best_modularity:
          best_modularity = modularity
          best_centers = centers

    // 6. Save results (time slice, centers, modularity)
    write_row_to_csv(
       data = [time_slice, metrics, best_centers_ID]
    )
```

## 3. Experiments

### 3.1. Benchmark experiments

#### 3.1.1. Datasets

To evaluate the performance of the STALS, four artificially generated networks derived from the benchmark datasets (i.e., Birthdeath, Expandcontract, Hide, and Mergesplit) are employed in this study. These synthetic networks were created by extending the static Lancichinetti-Fortunato

Radicchi (LFR) benchmark (Lancichinetti and Fortunato 2009) at five snapshots, allowing for the simulation of various community evolution patterns over time. For each snapshot, the true labels of communities are available. The four networks share common properties: an average node degree of 20, a maximum node degree of 40, and a community overlap parameter of 0.2. These synthetic networks were specifically designed to capture all possible community evolution scenarios.

*3.1.2. Baseline methods*

Six state-of-the-art approaches with different similarity measurements are set as the baselines compared with the STALS. Different from our framework, these six methods represent the pipeline of most existing community detection approaches. They initiate by scrutinizing the static communities and then adopt different similarity metrics to measure the evolution process of communities across time:

(1) Heuristic threshold-based method (Greene et al. 2010) first detected static communities at each timestamp, and then matched the detected communities with each dynamic network's front community. The matching process was evaluated by the Jaccard coefficient:

$$sim\left(C_{t_i}, C_{t_j}\right) = \frac{|C_{t_i} \cap C_{t_j}|}{|C_{t_i} \cup C_{t_j}|}, \quad (4)$$

where $C_{t_i}$ and $C_{t_j}$ represent two detected communities at the timestamp $t_i$ and $t_j$, respectively. The symbol || represents the member numbers. $C_{t_i}$ and $C_{t_j}$ are considered as a matched community if $sim(C_{t_i}, C_{t_j})$ is larger than a threshold k. Experiments demonstrated that the optimal k was 0.1.

(2) Text mining threshold-based method (Takaffoli et al. 2011) adopted a different similarity measurement to identify critical events in both consecutive and non-consecutive timestamps:

$$sim\left(C_{t_i}, C_{t_j}\right) = \begin{cases} \frac{|C_{t_i} \cap C_{t_j}|}{max\left(|C_{t_i}|, |C_{t_j}|\right)}, & if \frac{|C_{t_i} \cap C_{t_j}|}{max\left(|C_{t_i}|, |C_{t_j}|\right)} \geq k \\ 0, & otherwise \end{cases} \quad (5)$$

Experiments illustrated that the optimal threshold k was equal to 0.3.

(3) Group evolution discovery (GED) (Bródka et al. 2013) introduced a topological metric to optimize the defects of the former two methods that solely rely on a similarity metric to track community evolution:

$$sim\left(C_{t_i}, C_{t_j}\right) = \frac{|C_{t_i} \cap C_{t_j}|}{|C_{t_i}|} \cdot \frac{\sum_{C \in (C_{t_i} \cap C_{t_j})} NI_{C_{t_i}}(f_i)}{\sum_{C \in (C_{t_i})} NI_{C_{t_i}}(f_i)}, \tag{6}$$

where $NI_{C_{t_i}}(f_i)$ represents the weight of the node $f_i$ within the community $C_{t_i}$, which can be measured in various centrality metrics, e.g., betweenness centrality and degree centrality. Experiments showed that the optimal k was 0.1.

(4) The transition probability vector-based method (Tajeuna et al. 2015) utilized transition probability vectors to represent the degree of shared nodes between communities at different timestamps:

$$sim\left(C_{t_i}, C_{t_j}\right) = \begin{cases} \sum_{f_i}^{N_c} 2\frac{p_{C_{t_i},f_i} \cdot p_{C_{t_j},f_i}}{p_{C_{t_i},f_i} + p_{C_{t_j},f_i}}, & if \sum_{f_i}^{N_c} 2\frac{p_{C_{t_i},f_i} \cdot p_{C_{t_j},f_i}}{p_{C_{t_i},f_i} + p_{C_{t_j},f_i}} > k \\ 0, & otherwise \end{cases} \tag{7}$$

where $N_c$ is the total number of communities within the dynamic networks. $p_{C_{t_i},f_i}$ and $p_{C_{t_j},f_i}$ represent the components of the transition probability vectors $v_{C_{t_i}}$ and $v_{C_{t_j}}$. The threshold is automatically determined as the intersection point of two Gamma curves derived from similarity values between transition probability vectors.

(5) Identification of community evolution by mapping (ICEM) (Mohammadmosaferi et al. 2020) utilized a harsh-map to map the nodes in communities into a pair, which consists of a timestamp and a community index. The similarity between the two communities was defined as:

$$sim\left(C_{t_i}, C_{t_j}\right) = \frac{|C_{t_i} \cap C_{t_j}|}{|C_{t_i}|}, \tag{8}$$

$$sim\left(C_{t_j}, C_{t_i}\right) = \frac{|C_{t_i} \cap C_{t_j}|}{|C_{t_j}|}, \tag{9}$$

where $t_i < t_j$. Communities $C_{t_i}$ and $C_{t_j}$ are considered partially similar if $sim(C_{t_i}, C_{t_j}) > k_1$ and $sim(C_{t_j}, C_{t_i}) > k_1$. And they are very similar if $sim(C_{t_i}, C_{t_j}) > k_2$. Experiments proved the optimal $k_1$ was 0.1 and the optimal $k_2$ was 0.5.

(6) Overlap coefficient-based method (Mazza et al. 2023) adopted an overlap coefficient to measure the similarity between communities:

$$sim\left(C_{t_i}, C_{t_j}\right) = \frac{|C_{t_i} \cap C_{t_j}|}{min\left(|C_{t_i}|, |C_{t_i}|\right)}, \tag{10}$$

*3.1.3. Evaluation results*

The community detection performance was quantitatively assessed using Normalized Mutual Information (NMI), which provides a rigorous measure of similarity between detected and ground-truth communities:

$$NMI\ (X,Y) = \frac{2 * I(X,Y)}{H(X) + H(Y)}, \tag{11}$$

where $H(X)$ and $H(Y)$ represent the entropy of the detected communities and the ground-truth communities, respectively. $I(X,Y)$ represents the mutual information of X and Y. The closer the NMI value is to 1, the more accurate the community detection.

The comparative analysis in **Figure 3** reveals the NMI of our proposed method with the six baselines. Notably, our proposed method secures high NMI values across all datasets, demonstrating remarkable accuracy in community identification. While Greene and Mazza's methods occasionally obtain marginally better performance (difference less than 0.05), this minor variation likely stems from their specific optimization for unique network characteristics rather than representing a general advantage. The superior performance of STALS, Mazza, Greene, ICEM, and Tajeuna on the first three datasets suggests these methods effectively capture essential community structures, whereas the comparatively weaker results from GED and Tajaffoli may indicate limitations in their adaptability to evolving networks. A particularly noteworthy finding is STALS's upward performance trend with increasing snapshots, contrasting with most methods' stagnant or declining trends. This strongly suggests that our method's adaptive mechanism becomes increasingly effective as more temporal information becomes available. However, all methods faced challenges with the Mergesplit dataset, exhibiting both their worst performances and a general downward trend. This degradation likely reflects the difficulty in tracking communities, which are undergoing structural transformations caused by mergers and splits. Despite this, STALS maintains an NMI above 0.95, demonstrating exceptional robustness. These results collectively indicate that STALS achieves an effectiveness between sensitivity to community detection and resilience to structural noise, which stems from its ability to simultaneously consider multiple topological and temporal features.

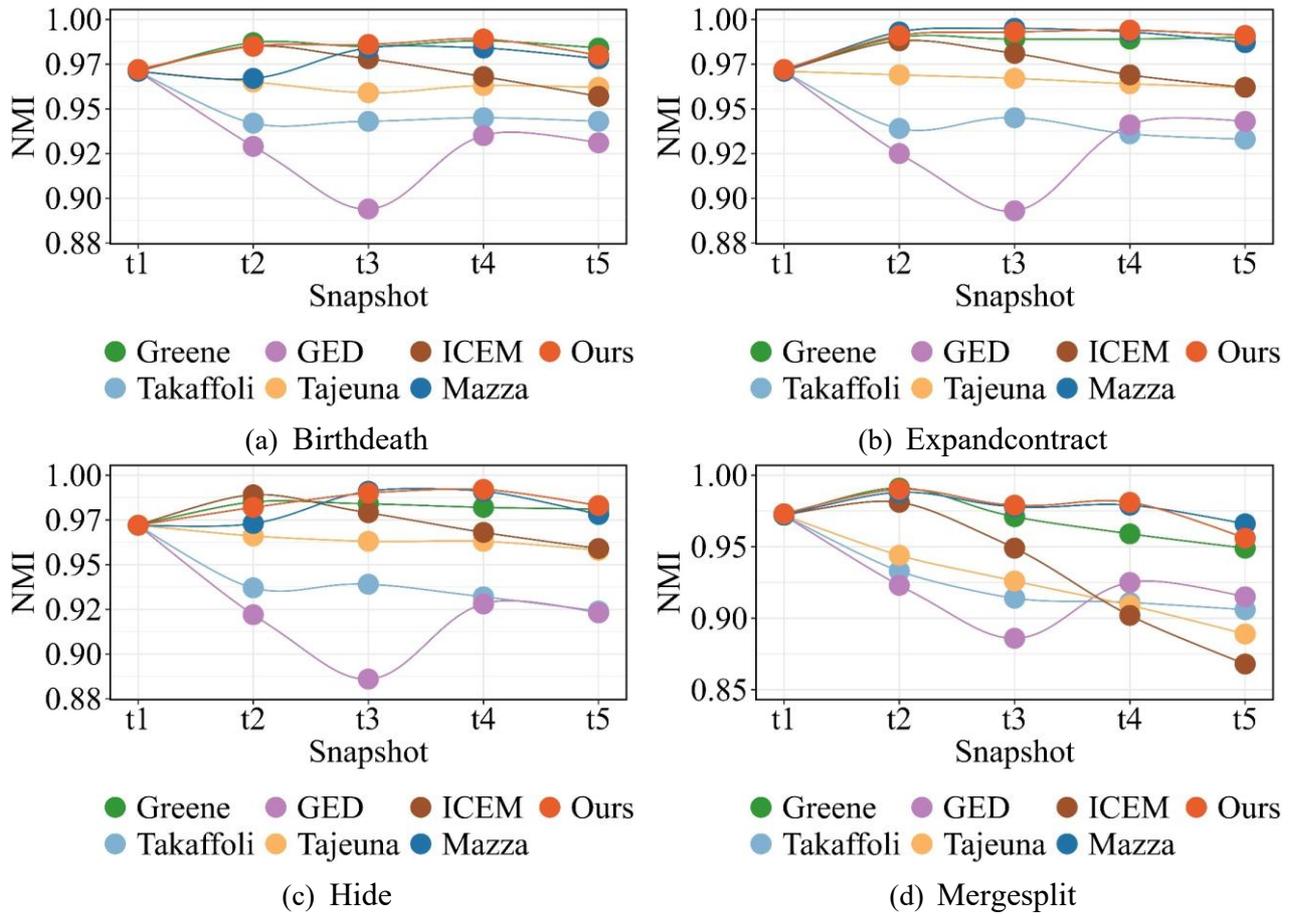

**Figure 3.** NMI of the proposed method and the six baseline methods on the four benchmark datasets.

*3.1.4. Execution efficiency*

**Tables 2-5** report the attributes of each dynamic network, including the community centers and running time of the STALS. The results reveal that the average execution time for the four dynamic network scenarios - Birthdeath, Expandcontract, Hide, and Mergesplit - is remarkably low, with values of 27.99s, 27.80s, 26.08s, and 28.77s, respectively. Notably, the maximum execution time does not exceed 30 seconds, reflecting the high efficiency of the STALS. To put this into perspective, our method outperforms the most efficient method (i.e., Mazza), which showed an average running time of about 31 seconds on a network with 15000 nodes at five snapshots. This is a significant improvement, especially considering the complexity of the networks being analyzed.

**Table 2** Execution efficiency on Birthdeath dataset.

| Timestamps | Nodes | Edges | Community centers | Running time |
| --- | --- | --- | --- | --- |
| 1 | 15000 | 71225 | 127 | 2321 ms |
| 2 | 14581 | 68539 | 123 | 1730 ms |
| 3 | 13997 | 65032 | 131 | 2103 ms |
| 4 | 13289 | 60997 | 114 | 1142 ms |

| 5 | 12781 | 58558 | 125 | 1101 ms |

Table 3 Execution efficiency on Expand dataset.

| Timestamps | Nodes | Edges | Community centers | Running time |
| --- | --- | --- | --- | --- |
| 1 | 15000 | 71225 | 127 | 2065 ms |
| 2 | 14807 | 69663 | 130 | 1686 ms |
| 3 | 14711 | 68305 | 137 | 1896 ms |
| 4 | 13289 | 60997 | 114 | 1142 ms |
| 5 | 14727 | 68106 | 141 | 1552 ms |

Table 4 Execution efficiency on Hide dataset.

| Timestamps | Nodes | Edges | Community centers | Running time |
| --- | --- | --- | --- | --- |
| 1 | 15000 | 71225 | 127 | 1844 ms |
| 2 | 13641 | 64239 | 111 | 1458 ms |
| 3 | 13584 | 63408 | 100 | 1520 ms |
| 4 | 13707 | 64019 | 113 | 1359 ms |
| 5 | 13633 | 63506 | 133 | 1645 ms |

Table 5 Execution efficiency on Mergesplit dataset.

| Timestamps | Nodes | Edges | Community centers | Running time |
| --- | --- | --- | --- | --- |
| 1 | 15000 | 71225 | 127 | 1844 ms |
| 2 | 15000 | 70290 | 120 | 1788 ms |
| 3 | 15000 | 69712 | 125 | 1701 ms |
| 4 | 15000 | 69485 | 149 | 2026 ms |
| 5 | 15000 | 69086 | 138 | 1272 ms |

*3.2. Case studies*

*3.2.1. Datasets*

(1) **Case 1**. NYC floating car data was downloaded from Uber Movement, covering the period from December 1, 2018, to December 31, 2018. The time interval is 1 hour. Each record contains the recording time, the road segment ID, and the average speed. The free-flow speed is acquired from the Uber Movement. Therefore, a 24-dimensional time series feature over one day can be obtained.

(2) **Case 2**. Shanghai taxi trajectory data was automatically collected by sensors, covering the period from April 1, 2016, to April 30, 2016. The data collection interval is 5 seconds. Each record contains the recording time, the road segment ID, and the actual speed. To compare with the NYC floating car data at different temporal resolutions, we aggregate the Shanghai taxi trajectory data into 5-minute intervals. In this way, a 288-dimensional time series feature over one day can be obtained.

(3) **Case 3**. Urumqi floating car data was collected by sensors, covering the period from October 1 to October 31, 2023. The data collection interval is 3 seconds. Each record contains the recording time, the road segment ID, and the actual speed. This kind of data was also aggregated into 5-minute intervals.

According to the data description of Uber Movement, the free-flow speed of a road segment equals the 15th percentile value of the actual speeds of all floating vehicles, with speeds sorted in descending order.

*3.2.2. Ablation study*

To systematically evaluate the performance of the STALS in tracking traffic congestion propagation in dynamic networks, we conducted a comprehensive ablation study examining the performance of different combinations of five features, i.e., node curvature (K), node degree (D), spatial proximity (S), TSI (T) time series, and the fast Fourier transform of the TSI (F). By utilizing the EW method for objective feature weighting, 12 distinct variants of adaptive adjacency matrices were generated: i.e., KS, KD, DS, KF, SK, DF, KDS, KDF, KSF, DSF, KDST, and KDSF. These combinations allow us to isolate and quantify the relative importance of different combinations in modeling the complex spatiotemporal dependencies, while evaluating the performance achieved by integrating features in pairs, triplets, and full quadruplets.

To rigorously assess the efficacy of our proposed method in real-world scenarios, modularity is employed as the evaluation metric. NMI is not applicable in this case because the ground truth community labels are not available. The selection of modularity is appropriate for this study as it is a mature index to effectively quantify community structure without predefined labels - a critical advantage given the lack of verified community partitions in real transportation networks. Formally, modularity (Q) measures the quality of a network partition by comparing the difference between the proportion of intra-community edges with the expected number of such edges in a random graph with an identical degree sequence (Newman and Girvan, 2004). The use of modularity aligns perfectly with our objective of identifying functionally congested communities in real urban settings where ground truth partitions are unavailable, which can be formulated as:

$$Q = \frac{1}{2m} \sum_{xy} [A_{ij} - \frac{k_i k_j}{2m}] \delta(C_i, C_j), \tag{12}$$

where $m$ is the total number of edges in the network. $A_{ij}$ represents the weight of the edge between nodes $i$ and $j$. $k_i$ and $k_j$ are the sum of the weights of the edges attached to nodes $i$ and $j$. $\delta(C_i, C_j) = 1$ if $i$ and $j$ belong to the same community (i.e., $C_i = C_j$). Otherwise, $\delta(C_i, C_j) = 0$. The range of Q is $[-1/2, 1]$. The higher the value of $Q$, the detected community quality is better.

The experiment results show distinct differences based on the 12 feature variants (**Figure 4**). First of all, KDSF achieves superior modularity on three datasets ($Q_{KDSF} > Q_{DSF} > Q_{KDF} > Q_{KSF} > Q_{KDST} > Q_{KDS}$), although Urumqi exhibits a slightly different ranking ($Q_{KDSF} > Q_{KDF} > Q_{DSF} > Q_{KSF} > Q_{KDST} > Q_{KDS}$) when integrating features in triplets and full quadruplets. The consistent superiority of KDSF confirms our hypothesis that a comprehensive combination of topological properties (K, D, S) with frequency-domain temporal features (F) yields optimal community detection, as it captures both structural relationships and mobility behaviors. Particularly, the striking performance between KDSF and KDST (with modularity improvement of 0.3, 0.28, and 0.24 for NYC, Shanghai, and Urumqi, respectively) highlights the advantage of frequency-domain transformation, suggesting that FFT processing more effectively extracts the inherent dynamics that align with community boundaries compared to raw time-domain series (T). Second, a strong showing of degree-containing combinations (particularly DSF and KDF) underscores the node degree's critical role as the most influential topological feature, because it directly reflects a road segment's functional connectivity in the network. Interestingly, Urumqi's unique result (KDF outperforms DSF) reflects the greater geometric importance because of its monocentric structure. Third, features combined in pairs (KS, KD, DS) exhibit poor effects for the three cities, which indicates that while topological properties provide useful supplementary information, they are far inferior to F. Collectively, these findings suggest that adaptive feature weighting could potentially enhance performance, but this does not mean that simply incorporating more features will yield better, as illustrated by $Q_{KDST} < Q_{KSF}$. The results provide empirical evidence that effective community detection requires both multi-scale topological analysis and sophisticated semantic information, instead of feature inclusion without discrimination.

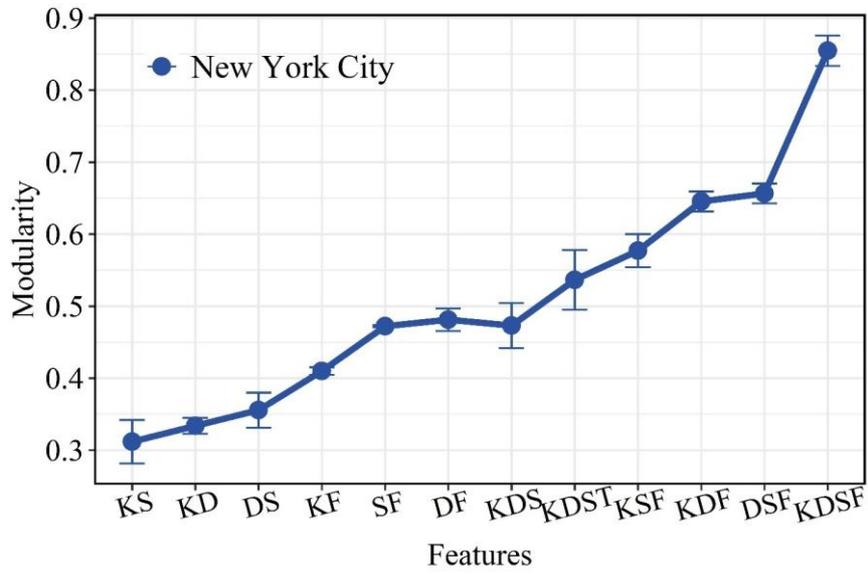

(a) New York City

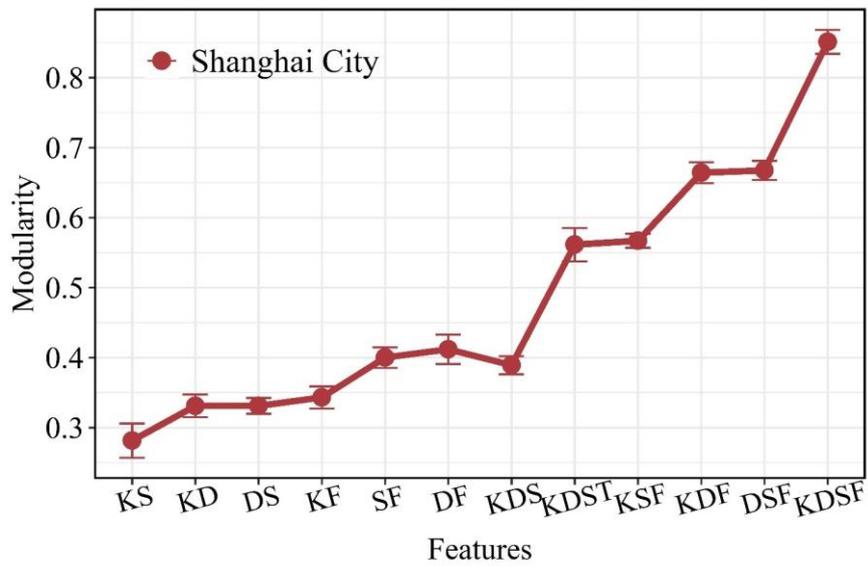

(b) Shanghai

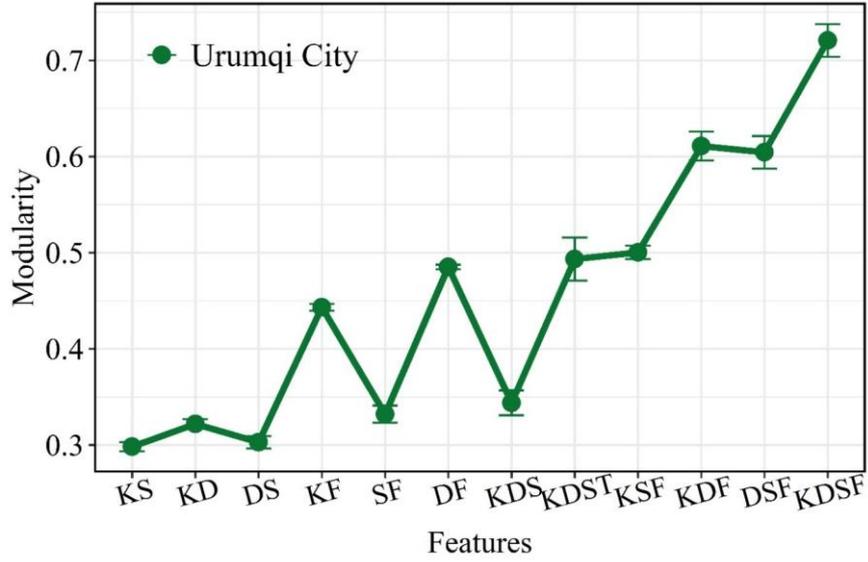

(c) Urumqi

**Figure 4.** Comparative analysis of modularity performance across 12 features combined in pairs, triplets, and full quadruplets for community detection.

*3.2.3. Propagation probability*

In this section, we aim to reveal some dynamics of the identified traffic congestion propagation bottlenecks by quantifying their propagation probability, which can be quantified as:

$$P_{C_k \rightarrow C_{k'}} = \frac{N_{C_k} \cap N_{C_{k'}}}{N} \quad (13)$$

where the $N_{C_k}$ and $N_{C_{k'}}$ represent the number of bottlenecks over different periods $C_k$ and $C_{k'}$, respectively. $k$ represnts the bottleneck level. $N$ is the number of total congestion bottlenecks.

The visualization in **Figures 5-7** reveals the transfer from high-grade bottlenecks to low-grade ones. Specifically, the outer circles in the diagrams represent bottlenecks at different levels, and the inner circles are cut into unequal parts, proportional to the chord thickness $P_{C_k \rightarrow C_{k'}}$. The comparative analysis reveals three distinct characteristics of bottleneck transmission patterns (**Table 6**). Our key findings include: (1) high-grade bottlenecks demonstrate significant stability. High-effect bottleneck categories (C2 and C3) exhibit a higher probability of self-shifting (30%~40%). This shift stems from the stability brought by commuter-related congestion, although temporary congestion in commercial areas and transportation hubs can introduce variability. However, we observe this stability is temporally modulated, with commercial districts showing greater weekend variability due to shifted travel purposes. (2) The directional asymmetry exists in bottleneck degradation patterns. High-grade bottlenecks are mostly transferred to low-grade categories, while reverse upgrades from low-grade to high-grade bottlenecks occur less frequently (lower than 15%). This suggests that congestion mitigation follows different mechanisms from congestion formation. (3) Traffic congestion bottleneck transmission exhibits spatial heterogeneity. The high-level bottleneck transmissions in Urumqi display concentrated spatial clustering, contrasting with the more dispersed distribution patterns observed in NYC and Shanghai, where C1 bottlenecks demonstrate connectivity across more categories.

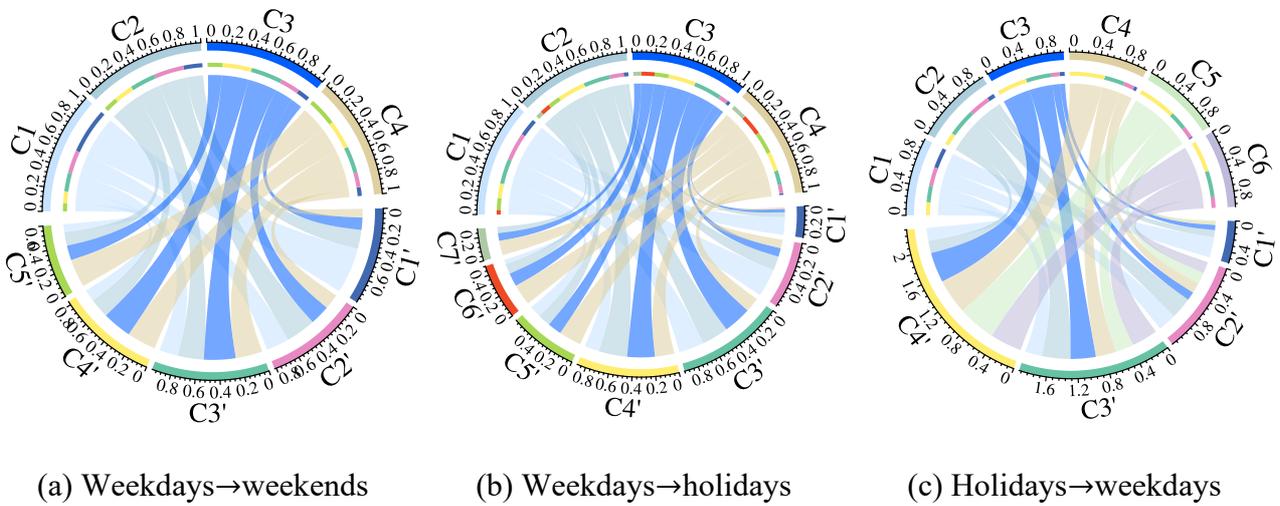

(a) Weekdays→weekends  (b) Weekdays→holidays  (c) Holidays→weekdays

**Figure 5.** Transitions of different grades of congestion bottleneck between different day types for NYC: C1-C6 (weekdays, holidays)→C1'-C7'(weekends, holidays, weekdays).

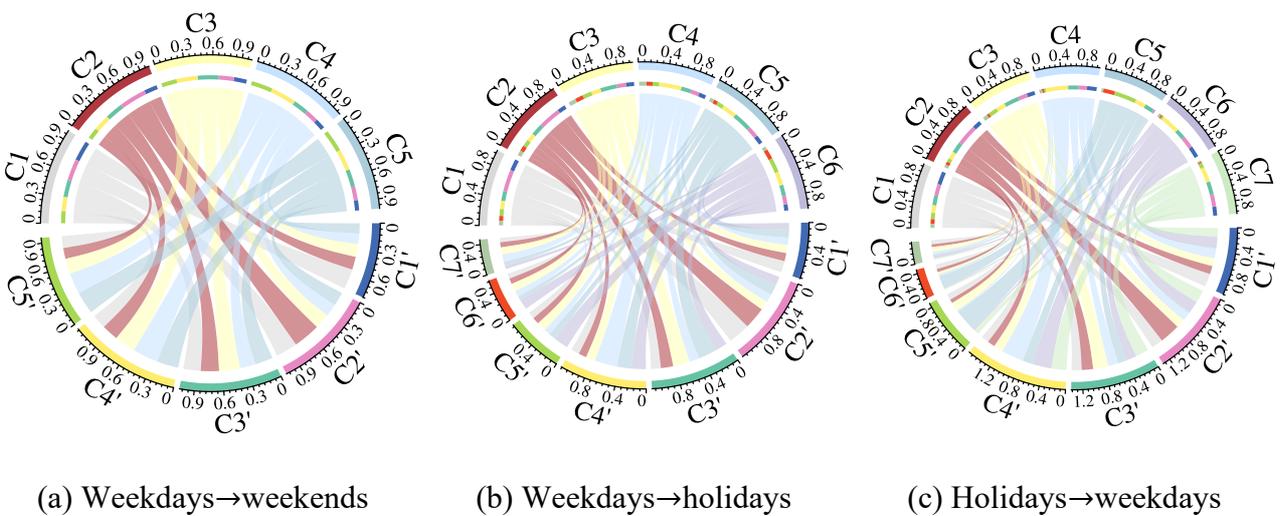

(a) Weekdays→weekends  (b) Weekdays→holidays  (c) Holidays→weekdays

**Figure 6.** Transitions of different grades of congestion bottleneck between different day types for Shanghai: C1-C7 (weekdays, holidays)→C1'-C7' (weekends, holidays, weekdays).

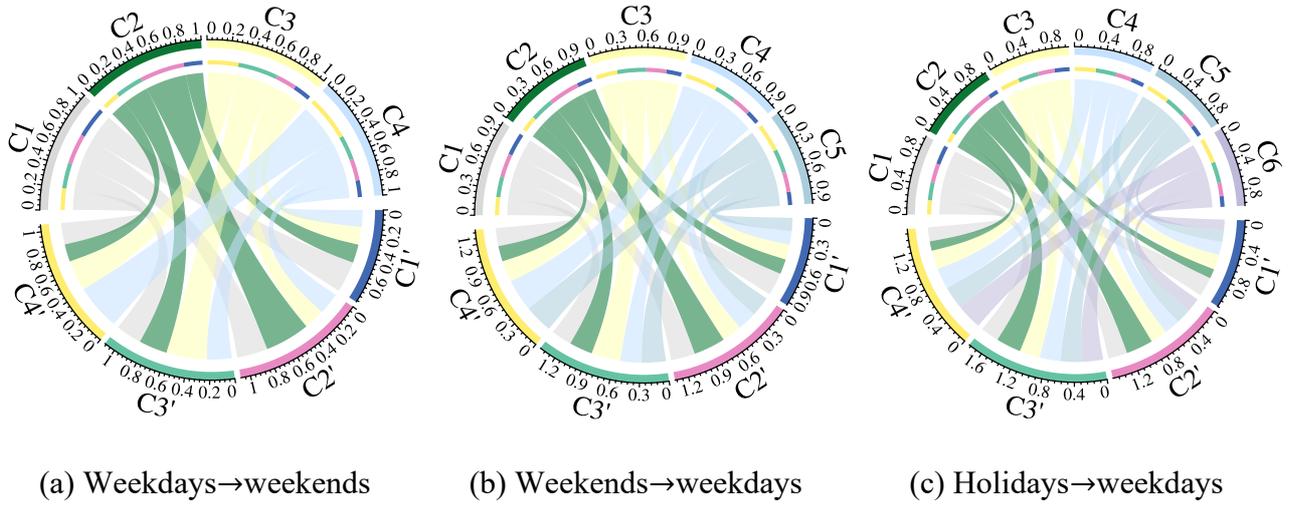

(a) Weekdays→weekends  (b) Weekends→weekdays  (c) Holidays→weekdays

**Figure 7.** Transitions of different grades of congestion bottleneck between different day types for Urumqi: C1-C6 (weekdays, holidays)→C1'-C4' (weekends, holidays, weekdays).

**Table 6** A summary of congestion bottleneck propagation characteristics.

| | Stability of high-grade bottleneck | Directional difference of bottleneck decay | Centrality difference of bottleneck distribution |
|---|---|---|---|
| NYC | $P_{C_3 \to C_{3'}} \approx 30\%$ | $P_{C_4 \to C_{1'}} \approx 7.86\%$<br>$P_{C_5 \to C_{1'}} \approx 2.05\%$<br>$P_{C_6 \to C_{1'}} \approx 0.00\%$ | $[C_1, C_2, C_3, C_4, C_5, C_6] \to$<br>$[C_{1'}, C_{2'}, C_{3'}, C_{4'}, C_{5'}, C_{6'}, C_{7'}]$ |
| Shanghai | $P_{C_2 \to C_{2'}} \approx 32\%$ | $P_{C_5 \to C_{1'}} \approx 8.57\%$<br>$P_{C_6 \to C_{1'}} \approx 14.54\%$<br>$P_{C_7 \to C_{1'}} \approx 11.97\%$ | $[C_1, C_2, C_3, C_4, C_5, C_6, C_7] \to$<br>$[C_{1'}, C_{2'}, C_{3'}, C_{4'}, C_{5'}, C_{6'}, C_{7'}]$ |
| Urumqi | $P_{C_2 \to C_{2'}} \approx 37\%$ | $P_{C_5 \to C_{1'}} \approx 12.33\%$<br>$P_{C_6 \to C_{1'}} \approx 14.24\%$ | $[C_1, C_2, C_3, C_4, C_5, C_6] \to$<br>$[C_{1'}, C_{2'}, C_{3'}, C_{4'}]$ |

*3.2.4. Propagation pathway*

Through STALS-based traffic congestion propagation tracking, **Figures 8-10** reveal the evolution of distinct congestion bottlenecks across New York City, Shanghai, and Urumqi over four specific periods. By focusing on two critical bottleneck zones (A and B) in each city, we uncover some latent dynamics that challenge conventional urban mobility assumptions and policy blind spots that shape congestion propagation. In New York City, polycentricity-induced congestion bottlenecks are distributed across multiple boroughs (**Figure 8**), which exposes a critical limitation in polycentric urban structures: the Brooklyn-Queens border (area A) experiences prolonged congestion compared with Manhattan's sharp 3-hour peak (16:00-19:00), indicating heavy reliance on limited transfer corridors of suburban commuters. This aligns with previous critiques of NYC's "last-mile" connectivity gaps, where subway and bus networks fail to efficiently distribute commuters beyond major hubs. It can also be noted that area B shows residual congestion bottlenecks near entertainment districts, implicating tourist mobility as an emerging bottleneck driver - a factor often overshadowed by commuter-centric models. The key implication is that polycentric urban structures do not eliminate congestion but rather prolong congestion propagation bottlenecks at critical interborough chokepoints. These findings challenge conventional assumptions about decentralization as a good congestion solution and suggest that effective management in NYC requires cross-borough connectivity and multimodal integration at key transfer points, rather than simply relying on spatial dispersion of urban functions.

Shanghai's congestion bottleneck distribution reveals a complex spatiotemporal evolution shaped by its grid-like road networks. While **Figure 9** shows regular progressive dispersion from morning to evening peaks, the downtown (area A) maintains remarkably uniform congestion distribution between 17:00 and 24:00. This contrasts with daytime's spatial concentration because (1) return trips typically involve multiple activities (shopping, dining, and social activities), while morning commutes are predominantly single purpose (home to work) with tighter time constraints; and (2) morning commuters concentrate on metro lines toward central business districts, but evening return trips dispers across private vehicles, ride-hailing, and public transit. Compared to urban core congestion, suburban roads (area B) along the arteries between the suburban rings and outer rings last longer as congestion bottlenecks, revealing structural deficiencies in suburban transportation systems. This persistent congestion phenomenon can be attributed to two aspects. First, the road network design fails to accommodate the complex travel patterns emerging from rapid suburban development, and is difficult to handle the reverse commuters returning from urban centers, and intra-suburban trips combining multiple purposes (work, education, and shopping). Second, the spatial dispersion of suburban trips' origins and destinations, combined with less synchronized activities, results in prolonged congestion bottlenecks. This finding demonstrates that suburban transportation systems cannot simply replicate urban solutions, but require redesigned strategies accounting for dispersed trip origins and bottlenecks.

The distribution of identified congestion bottlenecks in Urumqi presents a contrast to both NYC and Shanghai, revealing fundamental differences in monocentric mobility networks. As shown in **Figure 10**, the detected congestion bottlenecks remain intensely concentrated within the compact urban core, with Xinshi (area A) and Tianshan (area B) districts exhibiting the worst condition during evening peak hours (17:00-20:00). This spatiortemporal concetration directly results from Urumqi's radial transportation network and highly centralized job-housing distribution, where overe a half of employment remains concentrated within the city center. The monocentric structure produces several distinctive effects: first, the clearance time of congestion bottlenecks generated during peak periods is shorter than in NYC and Shanghai; and second, the congestion bottlenecks rarely extend beyond the core areas even during severe conditions, reflecting a localized mobility congestion challenge. This finding carries important implications that while polycentric cities distribute congestion bottlenecks across space, monocentric cities concentrate them both spatially and temporally. The evidence suggests that urban structure fundamentally determines the prolonged moderate delays in polycentric cities and shorter durations in mononcentric cities of the identified congestion propagation bottlenecks.

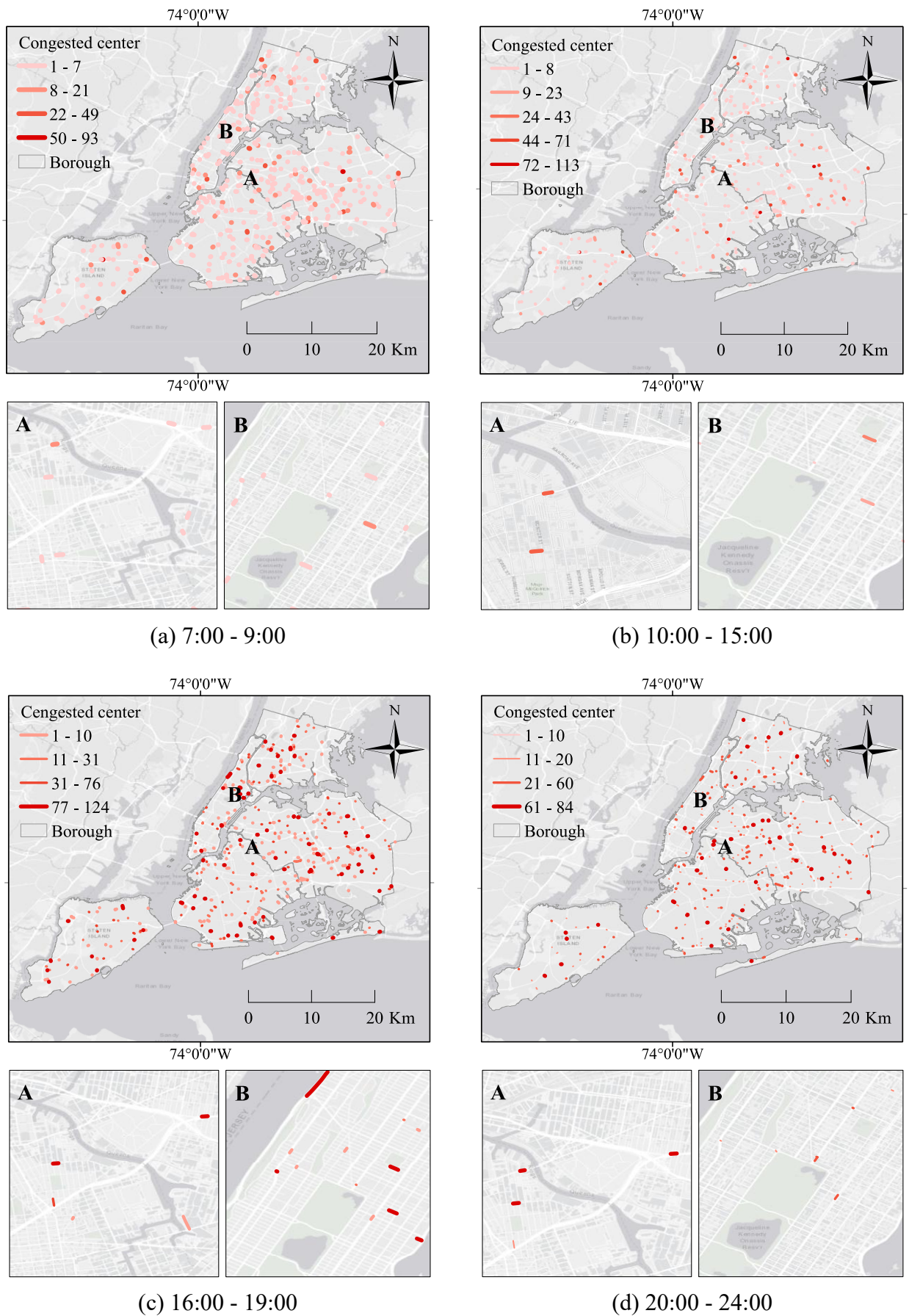

**Figure 8.** Propagation pathways over time in NYC.

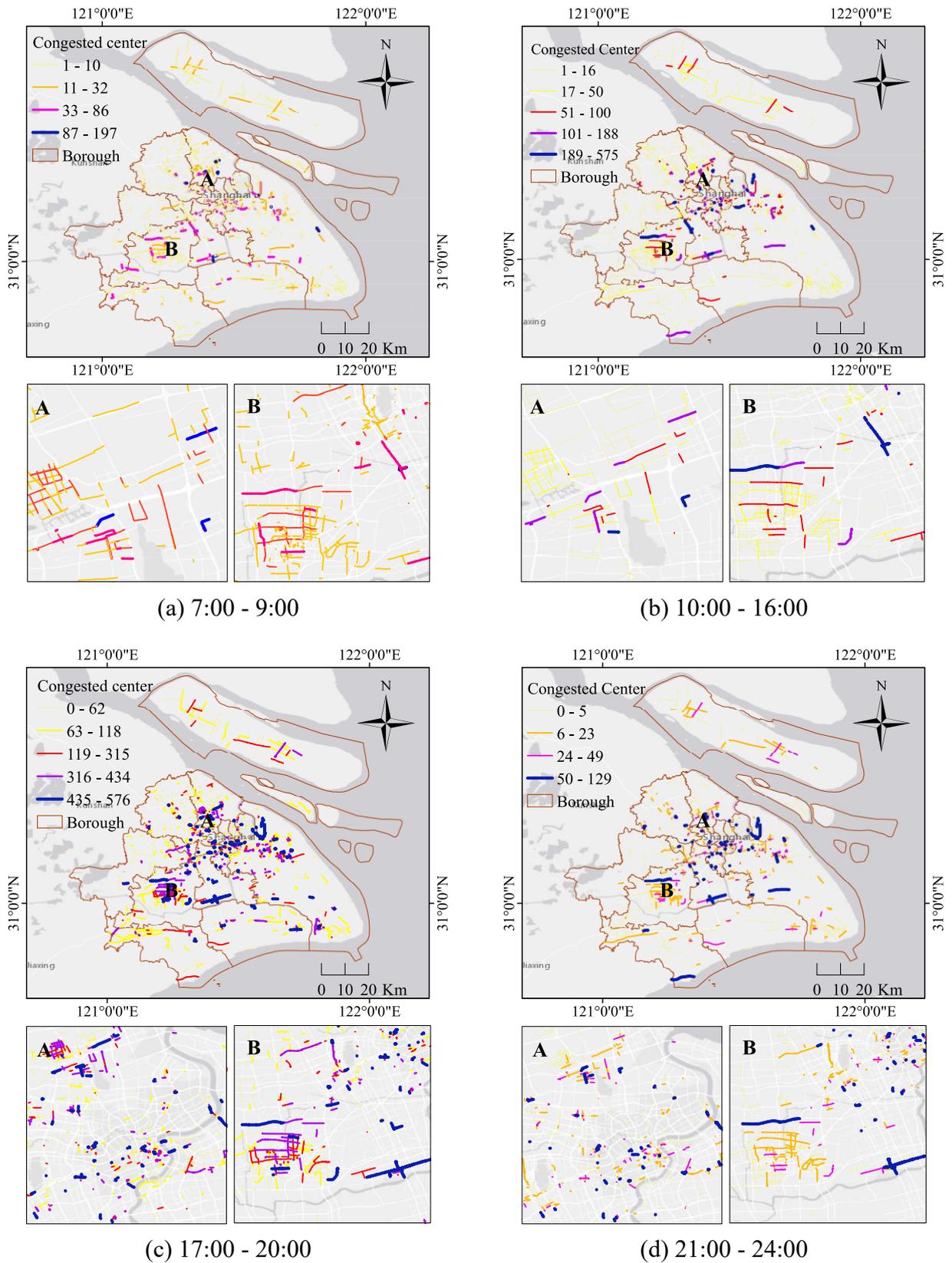

**Figure 9.** Propagation pathways over time in Shanghai.

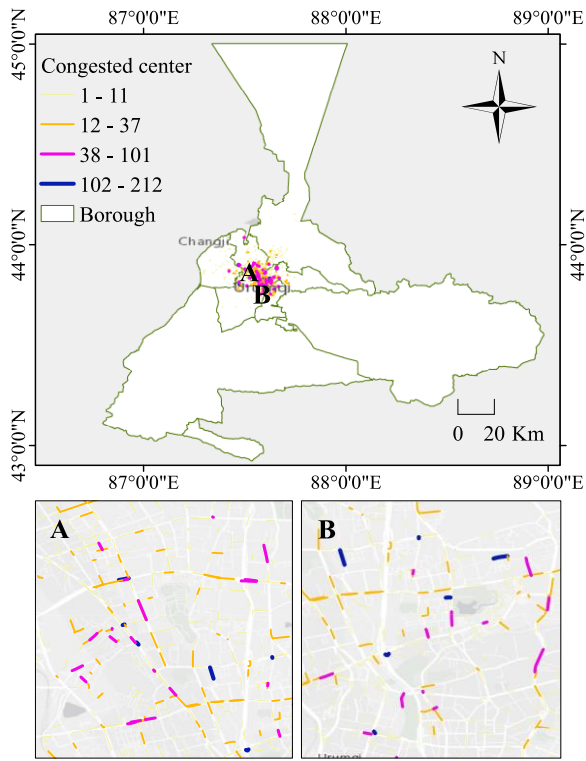
(a) 7:00 - 9:00

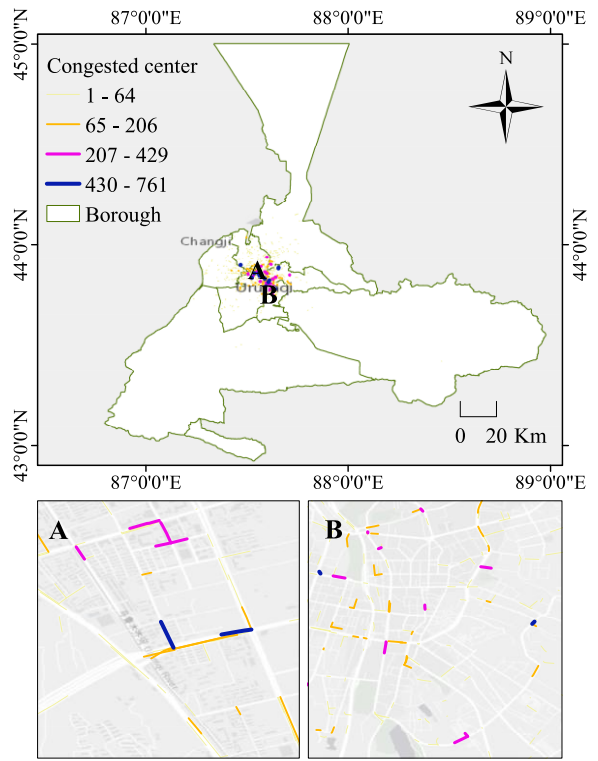
(b) 10:00 - 16:00

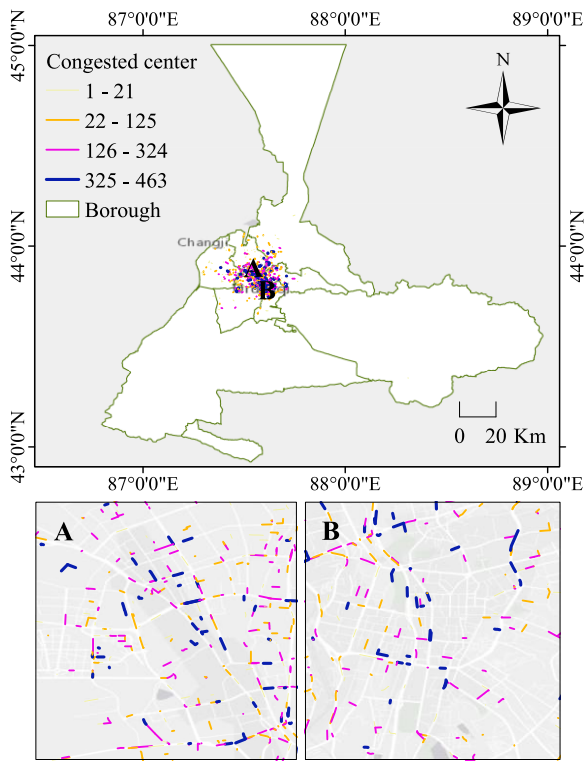
(c) 17:00 - 20:00

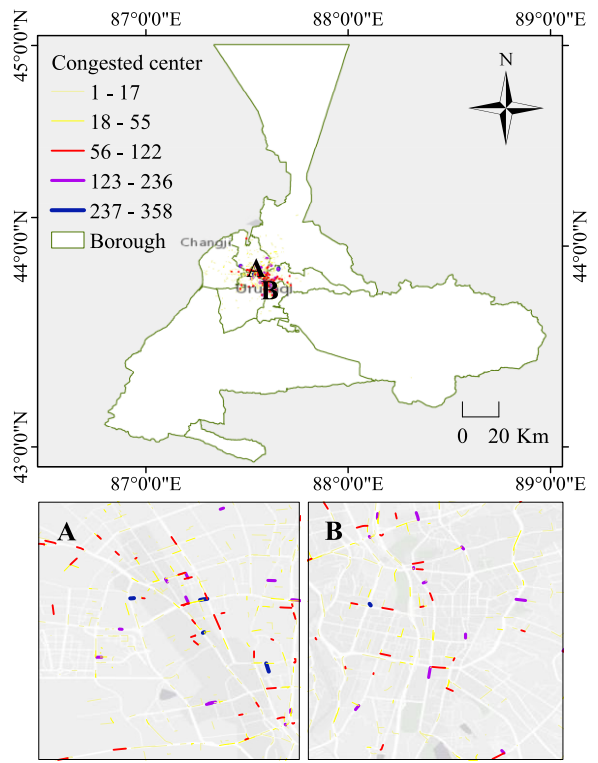
(d) 21:00 - 24:00

**Figure 10.** Propagation pathways over time in Urumqi.

## 4. Discussion

### *4.1. Feature importance*

The rationale for feature selection warrants further discussion. To achieve this, we analyzed the contribution of each factor by visualizing its effect when combined with others (F and T could not be jointly assessed due to their inherent relationship). As illustrated in **Figure 11**, the feature importance rankings - F>D>S>K>T for Shanghai and New York, and F>D>K>S>T for Urumqi - reveal several critical insights into their effectiveness in identifying traffic congestion propagation:

First, F consistently emerges as the most influential feature in three areas, which arises from its unique capability to decompose complex traffic states into two valuable frequency components: high-frequency fluctuations (characteristic of transient disturbances such as incident-induced congestion) and low-frequency periodic dynamics (e.g., recurring commuting rhythms). Conversely, the raw traffic state index (T) consistently ranks the lowest because it suffers from three key limitations that F effectively addresses: (1) as a purely time-dependent representation, it is difficult to distinguish random and persistent fluctuations from systematic congestion propagation; (2) The lack of frequency characteristics in T prevents the identification of recurring dynamics that often precede bottlenecks; (3) T is sensitive to noise and local variations, which frequently generate misleading congestion trends. In contrast, F provides inherent noise reduction through frequency-domain filtering while preserving the essential congestion variations. This explains why F achieves significantly better bottleneck identification.

Second, the degree of network nodes (D) maintains a stable second-place ranking across all cities, indicating its reliability as a topological metric for bottleneck identification. Despite this, its power exhibits spatial heterogeneity: while contributing approximately 40% performance improvement in Urumqi, it provides less than 30% enhancement in NYC and Shanghai. This stems from the fundamental differences in urban network topologies and routing flexibility: the hierarchical road network characterized by few redundant pathways and greater betweenness centrality variance in Urumqi shows stronger dependency on critical junctions; while highly interconnected network in

NYC and grid-like structure in Shanghai's grid provide more alternative paths that effectively distribute traffic loads across multiple local streets, thereby diminishing the power of individual node degrees. NYC and Shanghai's equitable topologies, with their flatter centrality distributions and higher route redundancy, enable congestion to bypass high-degree nodes. This finding reveals that degree centrality matters in hub-dominated networks, but its utility diminishes in more homogeneous or densely interconnected networks.

Third, while both curvature (K) and spatial distance (S) demonstrate moderate effectiveness across three cities, their relative performance reveals slight regionally dependent variations: S slightly outperforms K in NYC and Shanghai, whereas the opposite holds true in Urumqi. This can be explained by examining what each metric fundamentally captures: K quantifies the geometric complexity of road segments (sharp turns, winding paths), while S measures simple distance between locations. In NYC and Shanghai's systematically regular grid layouts, S proves more reliable because geometric variations are minimized through standardized road designs. Conversely, Urumqi is shaped by natural topography with sharp turns and irregular intersections, which results in the direct distance between points often failing to reflect actual travel paths. This finding implies how urban morphology fundamentally mediates the relationship between network structure and traffic dynamics: geometrically optimized cities are suggested to privilege distance-based metrics, while geometric-complexity measures become essential in topographically constrained urban cities.

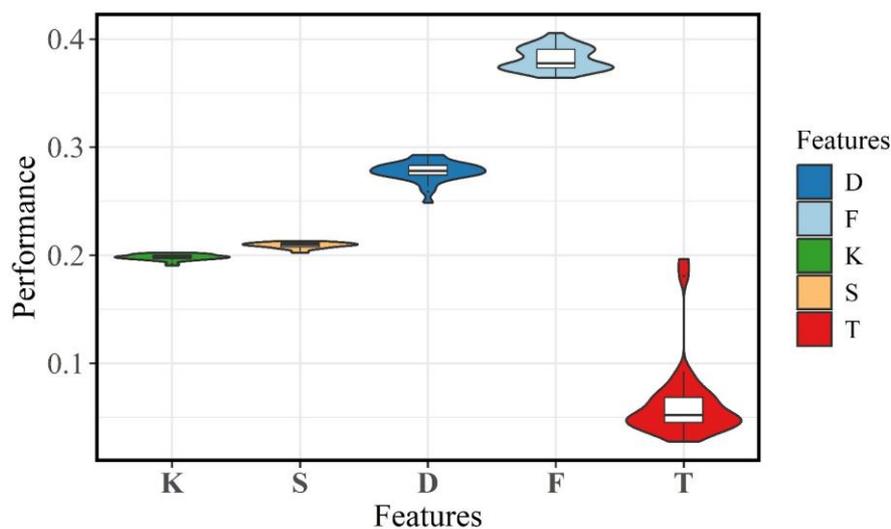

(a) New York City

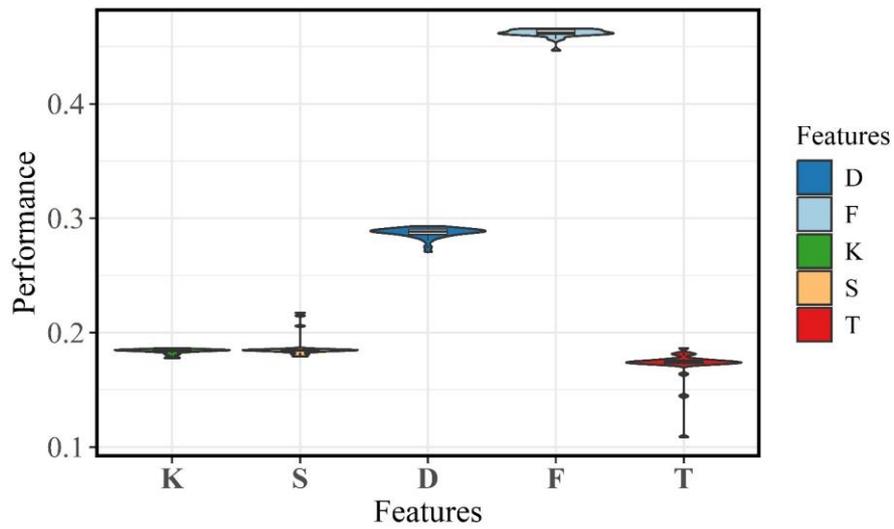

(b) Shanghai City

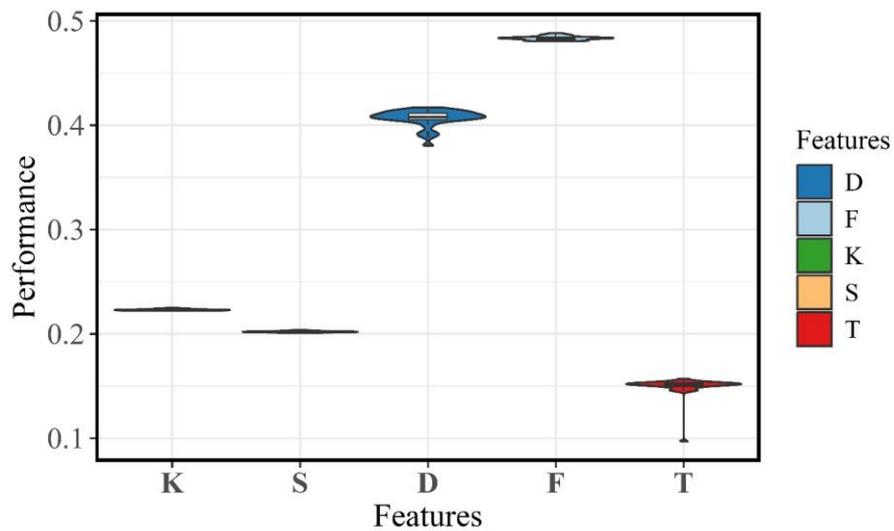

(c) Urumqi City

**Figure 11**. Performance of each feature, K, S, D, F, and T. F demonstrates consistent dominance across three areas, while the relative importance of D and S exhibits spatial heterogeneity.

## 4.2. STALS significance

The proposed spatiotemporal adaptive local search (STALS) method represents a methodological innovation in urban traffic propagation analysis by addressing three limitations in current congestion modeling approaches:

The first one is a dynamic congestion graph representation. To the best of our knowledge, predefined adjacency matrices are typically used in traffic congestion detection and forecasting tasks, particularly road connectivity-based matrices and distance-based matrices (Luan et al. 2022). However, empirical investigations reveal that such static representations suffer from a notable drawback - this pre-definition makes it difficult to capture the dynamic spatiotemporal relationship between congestion graphs, as they evolve over time (Jiang and Luo 2022). To address this issue, the STALS method utilizes an adaptive adjacency matrix to capture the spatiotemporal dependence, which considers both road network topology and traffic congestion states.

The second one is robust time-series processing. Current traffic propagation models frequently encounter performance degradation when handling time-series data of varying granularities, primarily due to their reliance on time-domain features that are inherently sensitive to data intervals. The STALS framework overcomes this limitation through the FFT of the traffic state index series, which decomposes the raw TSI series into distinct frequency bands that correspond to different congestion phenomena. This decomposition enables granularity-independent analysis because congestion dynamics manifest in specific frequency ranges.

Lastly, the concept of STALS borrows the idea of a clustering algorithm, which is essentially an unsupervised method that offers distinct advantages over conventional methods. Most notably, it introduces modularity - a community detection metric - to evaluate identification results when ground-truth labels for real-world traffic bottlenecks are unavailable. This stands in sharp contrast to traditional graph neural networks and other deep learning approaches for traffic congestion propagation detection, which typically require labeled data for both model training and performance evaluation. The unsupervised nature of STALS addresses a critical practical limitation, as obtaining reliable labels for traffic congestion bottlenecks often proves challenging due to the dynamic nature of urban traffic systems and the substantial costs associated with manual annotation.

### *4.3. Limitations*

This work has several limitations that warrant discussion. First, while the entropy weight (EW) method demonstrates reasonable adaptability in combining the four graph features, this approach may compromise the completeness of relative closeness measures between features. Future work should investigate more sophisticated fusion techniques that can better preserve the hierarchical relationships between primary and secondary features while maintaining adaptive weighting capabilities. Second, the current trajectory dataset shows limitations in terms of data coverage, which fails to account for external factors such as traffic accidents, weather conditions, and anomalous events. These unaccounted variables may affect the generalizability of our findings, suggesting the need for multimodal data integration in subsequent research. Lastly, it should be noted that our comparative analysis was conducted within the specific context of community detection frameworks. While this design choice aligns with our primary research objective - to investigate the understudied potential of community structures in identifying and tracking congestion propagation bottlenecks - it necessarily limits the scope of methodological comparisons.

## 5. Conclusion

This study presents STALS, a data-driven spatiotemporal modeling framework that tracks traffic congestion propagation in dynamic networks and advances urban sustainability. Our work makes three key contributions to sustainable urban mobility:

Firstly, our method demontrates a critical linkage between network community structures and traffic congestion propagation dynamics: (1) the community detection results empirically demonstrate that congestion centers naturally align with topological communities in dynamic urban road networks, as evidenced by high NMI (over 0.95 across snapshots) and good modularity performance (0.72-0.85 across NYC, Shanghai, Urumqi). This alignment occurs because the community boundaries often coincide with some physical infrastructure constraints, such as bridges and tunnels (Sun et al. 2014). (2) The observed congestion bottlenecks propagate along community-based pathways, demonstrating that community structures serve as conduits for congestion transmission. These suggest that community detection doesn't merely describe network topology, but actively reveals the congestion dynamics. This offers practical implications that urban planners should leverage GIS-based community detection to design resilient road networks, rather than relying on simple spatial clusters.

Secondly, the experimental results based on 12 feature variants combined in pairs, triplets, and full quadruplets, acquired from node curvature (K), degree (D), spatial proximity (S), traffic state index (T), and its frequency-domain transformation (F), demonstrate the importance and rationale for feature selection in practical traffic congestion prediction models. We conclude that (1) F secures the most significance regardless of spatial variation (NYC, Shanghai, Urumqi) or temporal resolution (1h, 5 min). This spatiotemporal-scale invariance suggests that F fundamentally captures the intrinsic congestion dynamics, and (2) Topological features (K, D, S) exhibit spatial heterogeneity that varies by urban structure. Node degree shows universal importance (30%-40% improvement) as a network centrality measure, but K and S display moderate performance (around 20%) and urban structure dependence. This suggests that effective traffic congestion modeling requires a context-aware feature integration strategy, instead of feature inclusion without discrimination.

Lastly, this study bridges unsupervised community detection techniques with traffic congestion propagation modeling, effectively addressing several practical challenges in existing approaches: (1) compared to the current supervised methods, STALS shows reasonable computational efficiency and maintains promising performance without ground truth information, showing potential for practical implementation on large real-world networks, and (2) by combining dynamic graph learning and Geo-driven analytics, STALS provides a scalable tool for spatial decision-making of congestion mitigation. In summary, this work contributes to traffic congestion propagation modeling by eliminating reliance on labeled training data, simplifying parameter requirements, and demonstrating robust performance in dynamic networks across diverse urban environments.

**Acknowledgments**

This work is supported by the National Natural Science Foundation of China under Grant 42171452.

**Declaration of Interest statement**

No potential conflict of interest was reported by the author(s).

**Author contributions**

Weihua. Huan, Wei Huang, and Xintao Liu conceptualized the study and theoretical framework. Weihua Huan carried out the mathematical derivations and drafted the manuscript. Kaizhen Tan participated in the data processing and mathematical derivations. Shoujun Jia and Shijun Lu contributed to data collection and result interpretation. All authors who participated in the manuscript revision approved the final version of the manuscript and agreed to be accountable for all aspects of the work.

**ORCID**

Weihua Huan. https://orcid.org/0009-0004-1347-4587

Wei Huang. https://orcid.org/0000-0001-8324-8877

**Generative artificial intelligence**

ChatGPT version 4o was used for language improvements.

**Notes on contributors**

**Weihua Huan** is currently a dual PhD student at Tongji University and Hong Kong Polytechnic University. Her research interests include GIS in transportation and the interaction between human activities and urban transportation.

**Kaizhen Tan** is an undergraduate student majoring in Information Systems at Tongji University. His research interests lie at the intersection of information systems and smart cities, particularly in GeoAI, spatio-temporal data mining, urban computing, and social computing.

**Xintao Liu** is currently an Associate Professor at the Department of Land Surveying and Geo-Informatics at The Hong Kong Polytechnic University. His research interests include GI services, GIS in Transportation, urban computing, and urban mobility and sustainability.

**Shoujun Jia** is currently a Postdoc researcher at the University of Innsbruck. He received a Ph.D. degree in surveying and geoinformatics from Tongji University, Shanghai, China, in 2023. His research interests include high-dimensional spatiotemporal data processing and 3D scene understanding applications.

**Shijun Lu** is currently an Associate Professor at Xinjiang University, and also a PhD student a Tongji University. His research interests include GIScience, human mobility, and social events.

**Wei Huang** is a Professor at Tongji University in Shanghai, China, and Adjunct Professor at Ryerson University, Toronto, Canada. His research interests lie at the intersection of GIScience, geography, computer science, and computational social science, focusing on using GIS, geospatial big data, and new sensing technologies to progress the understanding of the mechanisms of the interaction between human activities, urban transportation, and social events.

**Data availability**

The data and code that support the findings of this study are available on Figshare (Huan, 2025).

Data and Code DOI: https://doi.org/10.6084/m9.figshare.29501855.v1